\documentclass[%
 reprint,
 amsmath,amssymb,
 aps,
]{revtex4-2}
\usepackage{amsmath}
\usepackage{array}
\usepackage{booktabs}
\usepackage{multirow}
\usepackage{graphicx}
\usepackage{dcolumn}
\usepackage{braket}
\usepackage{textcomp}
\usepackage{tikz}
\usepackage{bm}

\usepackage{hyperref}
\usepackage{pifont}
\usepackage{bbding}
\usepackage{soul} 
\usepackage{color, xcolor} 
\sethlcolor{yellow} 
\usepackage{enumitem}
\newcommand{\mybullet}{\tikz \fill circle (1.5pt);}
\begin{document}
\preprint{APS/123-QED}

\title{Practical continuous-variable quantum key distribution using dynamic digital signal processing: security proof and experimental demonstration}

	\author{Lu Fan\textsuperscript{1}} \author{Zhengyu Li\textsuperscript{2}}\email{lizhengyu2@huawei.com}\author{Sheng Liu\textsuperscript{3}, Xuesong Xu\textsuperscript{1}, Tianyu Zhang\textsuperscript{1}}\author{Jiale Mi\textsuperscript{1}, Dong Wang\textsuperscript{3}, Dechao Zhang\textsuperscript{3}, Han Li\textsuperscript{3}, Song Yu\textsuperscript{1}}\author{Yichen Zhang\textsuperscript{1}}\email{zhangyc@bupt.edu.cn}
	
    \address{%
		\textsuperscript{1}State Key Laboratory of Information Photonics and Optical Communications, School of Electronic Engineering, Beijing University of Posts and Telecommunications, Beijing 100876, China\\ 
        \textsuperscript{2}Central Research Institute, 2012 Labs, Huawei Technologies Co., Ltd, Shenzhen 518129, Guangdong, China\\ \textsuperscript{3}Department of Fundamental Network Technology, China Mobile Research Institute, Beijing, China
	}%

\date{\today}

\begin{abstract}
Digital signal processing technology has paved the way for the realization of high-speed continuous-variable quantum key distribution systems.
However, existing security proofs are limited to static digital signal processing algorithms, while practical systems rely on dynamic multiple-input multiple-output algorithms to compensate for time-varying channel impairments.
Our analysis reveals that the conventional dynamic algorithm, due to its non-unitary nature, systematically underestimates the excess noise, which in turn leads to security issues and the generation of insecure keys.
To close this gap, we propose a secure algorithm model, mapping the dynamic algorithm to an equivalent physical optical model whose security can be rigorously assessed.
Simulations illustrate the algorithm's non-unitary property and provide a quantitative analysis of the excess noise underestimation caused by the conventional algorithm.
We further experimentally validate the necessity of the proposed modeling for dynamic digital signal processing, achieving a secret key rate of 14.4 Mbps based on estimated excess noise of 0.07 shot noise unit; whereas the conventional algorithm would have dangerously overestimated the key rate to 28.2 Mbps with noise of 0.008 shot noise unit.
This work provides the essential security framework for dynamic digital signal processing, overcoming a critical impediment for the development of high-performance continuous-variable quantum key distribution systems.
\end{abstract}

\maketitle


\section{\label{sec:level1}Introduction}
Quantum key distribution (QKD) \cite{bennett2014quantum} can provide unconditional security at the theoretical and protocol levels \cite{pirandola2020advances,xu2020secure}, making it an important direction of s2cientific research in the current complex information security situation. 
In addition to discrete-variable (DV) quantum key distribution, continuous-variable (CV) QKD achieves key transmission by encoding information in the canonical component of light. 
It has the advantages of high rate and high compatibility with existing classical optical communication systems \cite{zhang2024continuous,usenko2025continuous}, and attracts widespread attention.
After decades of development, the theoretical security analysis of CV-QKD has been well established \cite{leverrier2015composable, leverrier2017security,portmann2022security,pirandola2021composable}, and many high-performance systems have been reported \cite{jouguet2013experimental,zhang2019integrated,zhang2020long,jain2022practical,tian2022experimental}.

The third-generation CV-QKD system \cite{zhang2024continuous, guo2021toward} has evolved with the advancement of digital signal processing (DSP) technology, and its security proof for single-polarization system has been completed \cite{qi2015generating,soh2015self,chen2023continuous}.
The combination of DSP and optical chips further enhance the system capability by tens of GHz \cite{roumestan2024shaped,hajomer2024continuous_passive,pan2025high,bian2024continuous,hajomer2024continuous,wang2025high}, promoting high performance, high integration, and ease of use.
Due to the potential for double efficiency through simultaneously modulating and detecting independently in two orthogonal states, dual-polarization systems represent the next significant developmental trend.
This further promotes compatibility with coherent communications, paving the transmission capacity to 100G/200G even higher.
Moreover, the definition of networking, routing, and slicing based on the unified frame structure will facilitate the co-fiber transmission of quantum and classical signals.
However, the dynamic channel variations within the dual-polarization system constitute an issue that subsequent DSP must address.
Correspondingly, the dynamic multiple-input multiple (MIMO) algorithms \cite{yang2015fast,pan2023simple} is commonly used to cope with time-varying polarization impairments in practical dual-polarization channels, such as drift of the state of polarization (SOP), polarization dependent loss (PDL), phase rotation and so on.

The shot noise unit (SNU), serving as the measure of system excess noise, requires precise calibration within the system.
In general cases, the conventional MIMO (C-MIMO) algorithm under the presence of channel PDL is non-unitary, which alters the SNU and leads to excess noise misestimation via the covariance matrix.
In this paper, we proposed the quantum MIMO (Q-MIMO) model which incorporates trusted noise into the C-MIMO algorithm to compensate for the altered SNU.
\begin{figure*}[ht]  %
    \centering  
    \includegraphics[width=0.85\textwidth]{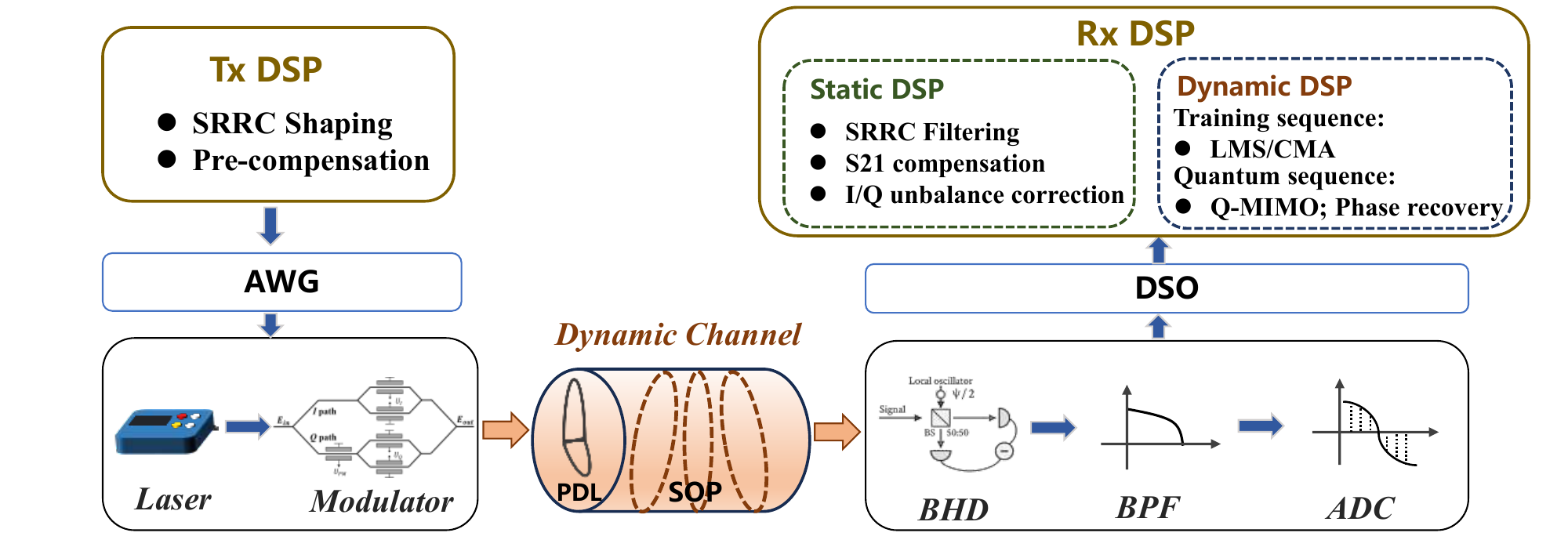}  
    \caption{The hardware architecture for the CV-QKD system and corresponding DSP compensation algorithms.
    SRRC, square root raised cosine; AWG, arbitrary waveform generato; BHD, balanced homodyne detector; BPF, band-pass filter; DSO, digital storage oscilloscope; ADC, analog to digital converter; LMS, least mean square; CMA, constant modulus algorithm.}  
    \label{fig:DSP_algo}  
\end{figure*}
We employ SVD decomposition as an analytical tool to demonstrate that Q-MIMO compensation can be mapped onto a series of physically describable matrices.
In the further equivalent EB model, Q-MIMO is modeled as a combination of passive linear optical networks and the beam splitter (BS)/phase insensitive amplifier (PIA) model whose security can be rigorously assessed. 
The simulation results illustrate the necessity of the Q-MIMO model in security analysis.
We then conduct the experimental verification on the 25.3 km fiber channel, achieving a secret key rate of 14.4 Mbps after applying the proposed Q-MIMO method.

The article is structured as follows, Part II starts with a brief introduction to the DSP algorithm used for CV-QKD. 
Then we conduct a theoretical analysis of the Q-MIMO model and show its necessity by simulations in Part III. 
The experimental validation is completed in Part IV.
We finally give the discussion and conclusion in Part V.

\section{\label{sec:level1}The brief introduction to the DSP algorithm for CV-QKD}
The adoption of DSP in digital CV-QKD systems offers the benefits of reduced hardware complexity and improved transmission capacity by suppressing noise. 
In digital systems, there are two types of sequence that serve different functions.
The slightly higher-power training sequence acquires channel response for damage compensation, and there are no constraints on its DSP algorithm. 
However, when processing the quantum sequence that contains key information, the security of the algorithms requires careful consideration.
The system hardware and common DSP algorithms are illustrated in FIG. \ref{fig:DSP_algo}.

Depending on the nature of the impairment, DSP algorithms are primarily categorized into two types.
Static equalization focuses on compensating for static hardware imperfections within individual detection channels, rather than the dynamic characteristics of the transmission link.
For example, SRRC matched filtering is a classical method for addressing ISI originating from limited bandwidth. 
For certain devices with non-flat frequency response, DSP can utilize the S21 parameters to design a filter with its inverse transfer function. 
Furthermore, the I/Q imbalance correction algorithm ensures that the I and Q signals are strictly orthogonal when entering the digital domain.
These algorithms rely on precise prior knowledge of the impairment model, which are specifically designed for single-detection channel.

Dynamic MIMO equalization is specifically designed to address varying fiber impairments that cause inter-channel crosstalk.
For example, due to residual birefringence within the optical fiber and external environmental disturbances, the SOP leads to severe crosstalk between the two orthogonal polarization channels.
The PDL causes signals in different polarization channels to experience unequal and varying attenuation.
The laser linewidth and residual frequency offset also induce a phase rotation of the canonical components.

\begin{figure}[t]  %
    \centering  
    \includegraphics[width=0.4\textwidth]{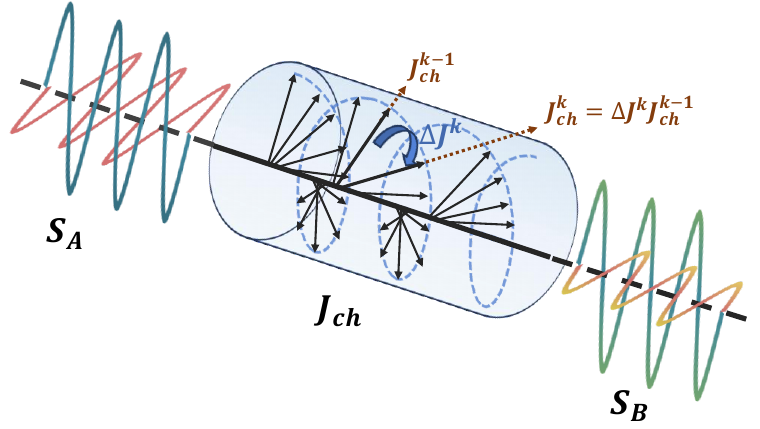}  
    \caption{The dynamic evolution model of generalized channel. The dual-polarization input signal $S_A$ is transformed into the output signal $S_B$ through the optical channel $J_{ch}$.
    The instantaneous channel response at point $k$ is recursively derived from the previous state $J_{ch}^k$
  and the incremental evolution matrix $\Delta J^k$ where 
$\Delta J^k$ accounts for polarization rotation, PDL and phase shift.}  
    \label{fig:Jch}  
\end{figure}

For a generalized channel, the dynamic evolution model characterized by Jones matrices $J_{ch}$ \cite{cui2018two} is illustrated in FIG.\ref{fig:Jch}.
The channel matrix $J_{ch}^k$ at time $k$ can be obtained by multiplying the matrix of the previous moment $J_{ch}^{k-1}$ by a small rotation matrix $\Delta J^k$, resulting in $J_{ch}^k=\Delta J^kJ_{ch}^{k-1}$.  
\begin{equation}
\label{J}
\Delta J^k = \begin{bmatrix}
    \cos(\Delta\theta_k)e^{j\phi_1} & -\sin(\Delta\theta_k)e^{j\phi_2} \\
    \sin(\Delta\theta_k)e^{-j\phi_2} & \cos(\Delta\theta_k)e^{-j\phi_1}
\end{bmatrix}\begin{bmatrix}
    1 & 0 \\
    0 & e^{-\eta_k}
\end{bmatrix}.
\end{equation}

The parameter $\Delta\theta_k$ is the polarization rotation angle, $\phi_1$ and $\phi_2$ are the phase angles. 
The second item originates from the PDL and phase difference, where $\eta_k$ is the coefficient associated with $PDL_{dB} = 20log_{10}e^{\eta_k}$. 
At any time point $k$, the total rotation matrix $J_{ch}^k$ is the result of the initial state $J_{ch}^0$ and the cumulative effect of incremental evolutions. 

The dual-polarization signal at sending can be described by $S_A =[ S_A^X,S_A^Y]^T$, and $S_A$ is actually the mean of the coherent state.
It undergoes through the channel response $J_{ch}$ and outputs $S_B$ as $S_B =[ S_B^X,S_B^Y]^T$.

Dynamic MIMO requires evaluating this channel response to recover signal.
This work focuses on single-tap MIMO algorithm, reserving complex multi-tap MIMO for subsequent research. 
Common MIMO algorithms include adaptive equalization \cite{widrow1988adaptive} and blind equalization \cite{godard2003self} . 
Their core principle involves training the transmission matrix $W$ to operate on the received signal $S_B$.
Ideally, the $W$ matrix should be the inverse of the channel response with $W=J_{ch}^{-1}$.

\begin{figure}[t]  %
    \centering  
    \includegraphics[width=0.3\textwidth]{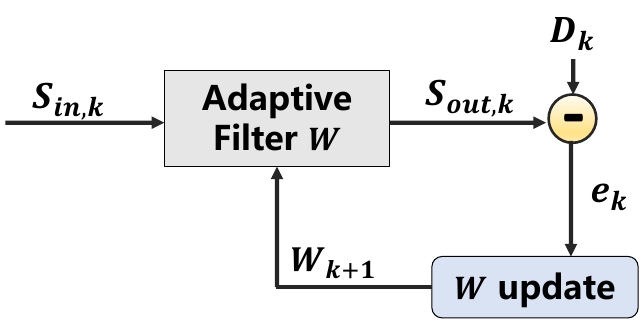}  
    \caption{The training process of LMS equalization algorithm. The filter matrix $W$ is dynamically updated based on the error  between the equalizer output and the original signal .}  
    \label{fig:LMS}  
\end{figure}
We will briefly introduce the training process of the $W$ matrix using the LMS algorithm as an example shown in FIG.\ref{fig:LMS}.
When the input $S_{in,k}$ is applied to the equalizer, its goal is to make output $S_{out,k}$ can best approximate the original transmitted signal $D_{k}$.
It will dynamically update the tap coefficients by iteratively minimizing the mean square value of the error signal. 
The update process is as follows:
\begin{itemize}[label=\small\mybullet, itemsep=-0.3em, topsep=0.1em]
\item Calculate the equalizer output: $S_{out,k}=W_k^HS_{in,k}$;
\item  Calculate the error: $e_{k}=D_{k}-S_{out,k}$;
\item  Update the coefficient matrix: $W_{k+1}=W_k+\mu S_{in,k}e^H_k$.
\end{itemize}
$\mu$ is the step size factor that balances convergence speed and steady-state error. 

The trained matrix compensates for attenuation and crosstalk in additive Gaussian white noise (AGWN) channels, with the compensation matrix $W$ being non-unitary in the presence of PDL.
Performing such non-unitary operation on quantum signals requires additional consideration, as it may compromise security.
\begin{figure*}[ht]  %
    \centering  
    \includegraphics[width=0.8\textwidth]{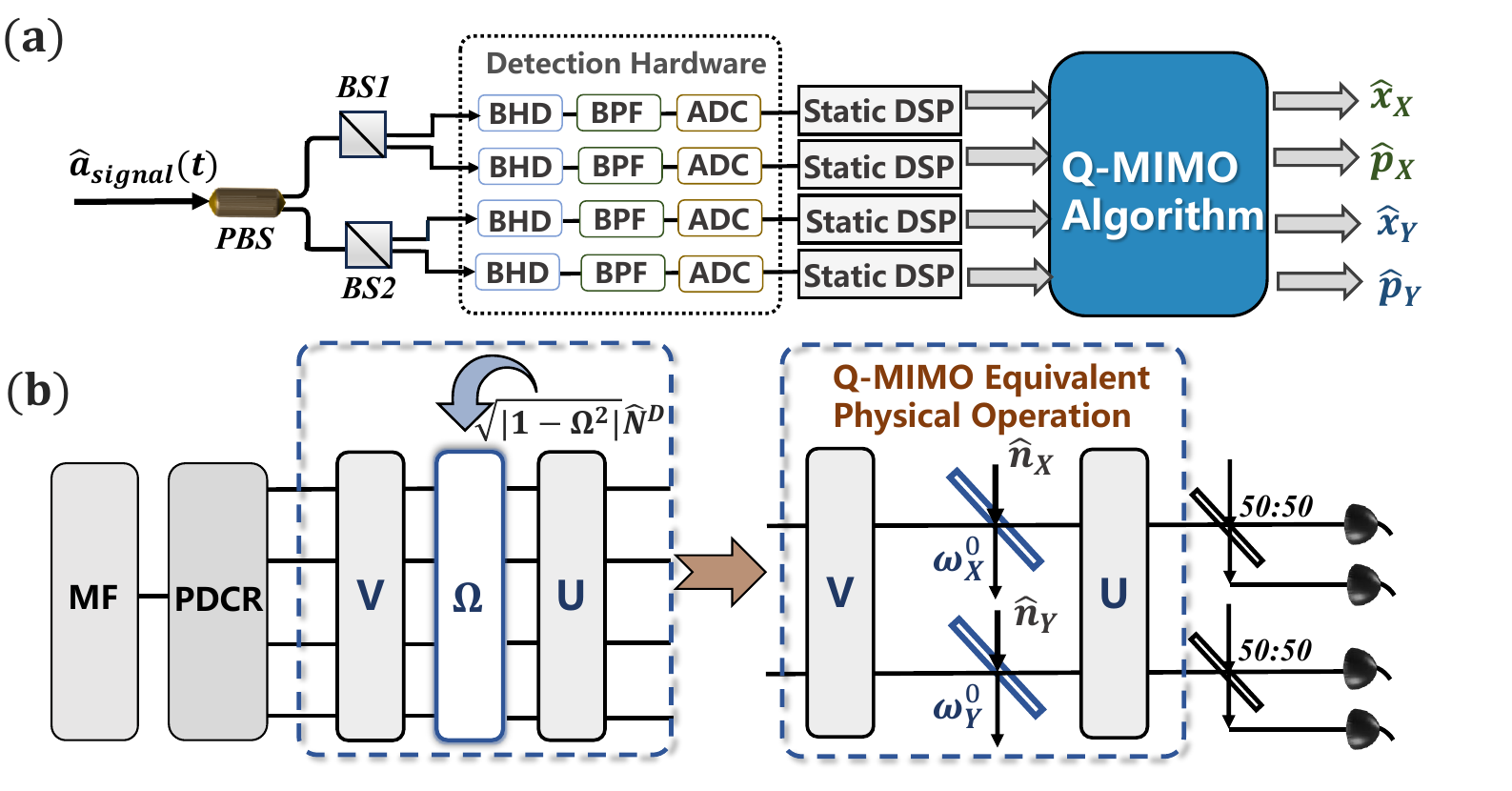}  
    \caption{(a) The practical architecture of hardware and softwore within the CV-QKD receiver. PBS, polarization beam splitter.
    (b) The full EB model of DSP algorithm for CV-QKD. In the first stage, the static DSP is modeled as MF. 
    In the second stage, the dynamic 1Q-MIMO is modeled as three equivalent physical operations.
    PDCR, polarization diversity coherent receiver. MF, mode-filter}  
    \label{fig:ICR}  
\end{figure*}

\section{\label{sec:level1}Q-MIMO Model for CV-QKD}
The structure of the receiver end in a dual-polarization CV-QKD system is shown in FIG.\ref{fig:ICR} (a). 
It includes practical detection hardware and software DSP algorithms.
The two-stage security framework for full DSP stack of CV-QKD is shown in (b).
In the first stage, the static DSP for single BHD detection channel is regarded as mode-filter through SNU normalization.
In the second stage, regarding the inter-channel crosstalk compensation, we add additional trusted noise into the C-MIMO output to map it to a physical operation suitable for quantum signals, named as the Q-MIMO model.
Through SVD decomposition, Q-MIMO can be represented as a combination of two unitary operations and a BS model.

\subsection{\label{sec:level1}Brief introduction of the security of static DSP}
Based on continuous-mode field theory \cite{blow1990continuum,raymer2020temporal,raymer1989temporal}, the generalized security proof for static DSP algorithms has been presented \cite{chen2023continuous}.
Inside the detection structure, the received quantum state with the annihilation operator $\hat{a}_{signal}(t)$ is divided into orthogonal polarization directions, denoted $\hat{a}_{X}(t)$ and $\hat{a}_{Y}(t)$.
Then, take the x-quadrature measurement as example, the output photocurrent of BHD after interference in the X-Pol can be directly written as
\begin{equation}
\label{f_LO}
\begin{split}
			\hat{f}_{LO}^X(t)={\int}d\tau[A_{LO}^X\xi_{LO}^X(\tau)exp(-i\omega_{LO}^X(\tau)+i\theta_{LO}^X)\hat{a}^{\dagger}_{X}(\tau)\\+A_{LO}^X{\xi_{LO}^X}^{*}(\tau)exp(i\omega_{LO}^X(\tau)-i\theta_{LO}^X)\hat{a}_{X}(\tau)]h^X(t-\tau).
        \end{split}
		\end{equation}
Here, the classical LO signal $a_{LO}^X(t)$ is expressed by the wavepacket function as $a^X_{LO}(t)=A^X_{LO}\xi^X_{LO}(t)exp(-i\omega_{LO}^{X}t+i\theta_{LO}^{X})$, and the finite bandwidth of the practical detector can be represented by the impulse response function $h^X(t)$.

Next, the output of the ADC is the integrated average of the data within $\Delta{t}$ near the sampling moment $t_j$, that is,
\begin{equation}
\label{D_tj}
			\hat{D}_{t_j}^X=\frac{1}{\Delta{t}}{\int}_{t_j}^{t_j+\Delta{t}}dt\hat{f}_{LO}^X(t).
		\end{equation}
        
It has already been mentioned that static DSP operation $f^{X,DSP}(N)$ involves a fixed set of $N$ filter tap coefficients, thereby yielding the output as
\begin{equation}
\label{D_tj,DSP}		\hat{D}_{t_j}^{X,DSP}=\Sigma^N_{i=1}f_{i}^{X,DSP}\hat{D}^X_{t_{j-k+i}}.
\end{equation}

Based on the orthogonal characteristics of wavepacket represented by RRC, the time-dependent wavepacket normalized by SNU is $\Xi_{DSP}^X(\tau)=\frac{1}{\sigma_{cal}}\Sigma^N_{i=1}f_i^{X,DSP}{\int}dt {h}^X(t-\tau)\xi_{LO}^X(\tau)exp(-i\omega_{LO}^X\tau)$ can be abstracted.

The result normalized by SNU is rewritten as
\begin{equation}
\begin{split}
	\hat{D}_{t_j}^{X,SNU}=\int{d\tau}\Xi_{DSP}^X(\tau)exp(i\theta_{LO}^X)\hat{a}_{\tau}^{\dagger}(\tau)+h.c..
            \end{split} 
		\end{equation}
        
It shows that the practical characteristics of the receiver hardware and the static DSP algorithm can be conceptualized as a pair of normalized temporal mode-filter and homodyne detection, whose output corresponds to the measurement operator with a specific wavepacket $\Xi_{DSP}^X(\tau)$. 
The optical phase $\theta_{LO}^X$ determines the output quadrature of the measurement operator in phase space.
When $\theta_{LO}^X=0$, the $x$-quadrature operator is $\hat{x}_{X}$ and $\theta_{LO}^X=\pi/2$ leads to the output operator $\hat{p}_{X}$ in the $p$-quadrature.
Similarly, there are also two quadratures in another polarization direction based on the difference in optical phase $\theta_{LO}^Y$, including $\hat{x}_{Y}$ and $\hat{p}_{Y}$.
Usually, static DSPs aim at compensating the hardware imperfections and matching to a same pulse shaping filter, and this leads to an almost same mode-filter for all BHD channels.

\subsection{\label{sec:level2}The security analysis of Q-MIMO algorithm}
Since MIMO algorithms compensate primarily for time-varying polarization-dependent impairments, we assume that the losses of quadrature operators for each polarization direction are balanced for simplicity.
In fact, through proper filtering or redefinition of the reference framework, different channels can achieve balance for mode-filter efficiency,  polarization-dependent efficiency, and detector-dependent detection efficiency.

Assume the complex representation of the dual-polarized data sent by Alice is $\alpha_A=[\alpha_X,\alpha_Y]^T$ when $\alpha$ refers to the mean of a coherent state.
The certain detection result after SNU normalization is 
\begin{equation}
{S}_{in}={\alpha}_{B}+\hat{N},
\end{equation}
where ${\alpha}_{B}=J_{ch}{\alpha}_{A}$ represents the mean value of received coherent state, and $\hat{N}$ the vacuum noise of homodyne detection whose variance is $\boldsymbol{1}$.

The transfer function trained by the MIMO algorithm can be described by a 2*2 matrix as
\begin{equation}
W=\begin{bmatrix}w_{XX}&w_{YX}\\w_{XY}&w_{YY}\end{bmatrix}.
\end{equation}
In the ideal equalization situation, $W$ is the inverse of $J_{ch}$, aiming to recover the output signal
\begin{equation}
{S}_{out}=W{\alpha}_{B}+W\hat{N}={\alpha}_{A}+\hat{N'}.
\end{equation}

In general cases, $W$ matrix is non-unitary, so that SNU is modified as
\begin{equation}
var(N')=\begin{bmatrix}|w_{XX}|^2+|w_{YX}|^2\\|w_{XY}|^2+|w_{YY}|^2\end{bmatrix}\neq \boldsymbol{1}. 
\end{equation}
If the MIMO output $S_{out}$ is used for covariance calculations, parameter estimation results will be misjudged.
\begin{itemize}[label=\mybullet, itemsep=-0.3em, topsep=0.1em, leftmargin=*]
\item
When $|w_{XX}|^2+|w_{YX}|^2> 1$, $Var(N'_X)$ in the X-Pol exceeds 1, leading to overestimation of excess noise that makes the system unusable. 
\item 
When $|w_{XX}|^2+|w_{YX}|^2<1$, $Var(N'_X)<1$, the estimated excess noise is underestimated, resulting in an underestimation of eavesdropping and compromising security.
\end{itemize}

It can be seen more clearly through the SVD decomposition of the W matrix, with the 
\begin{equation}
W = U\Omega V=U\begin{bmatrix}\omega_{X}&0\\0&\omega_{Y}\end{bmatrix}V.
\end{equation}
Both $U$ and $V$ are unitary matrices representing polarization rotation. 
$\Omega$ is a scaling matrix whose two values may differ due to the presence of channel PDL. Here, $\omega_{X/Y}\textgreater1$ indicates gain while $\omega_{X/Y}<1$ indicates attenuation.

We propose to add trusted noise in digital domain after $W$ compensation, in order to map the whole process to a equivalent physical operation.
\begin{itemize}[label=\mybullet, itemsep=-0.3em, topsep=0.1em, leftmargin=*]
\item 
\textbf{For gain:} The noise term $\sqrt{(\omega_{X/Y}^2-1)}\hat{N}^{D}$ is added, making the operation equivalent to a PIA.
\item 
\textbf{For attenuation:} The noise term $\sqrt{(1-\omega_{X/Y}^2)}\hat{N}^{D}$ is added, making the operation equivalent to a BS.
\end{itemize}

Here, $\hat{N}^{D}$ denotes Gaussian random numbers with a variance of 1.
Obviously, the output following the Q-MIMO model is written as
\begin{equation}
{S}_{out}^Q={\alpha}_{A}+U(\Omega V\hat{N}+\sqrt{|1-{\omega_{X/Y}}^2|}\hat{N}^{D}).
\end{equation}

Given the complexity of integrating a dynamic gain PIA model, the attenuation model is always more practical for Q-MIMO with polarization recovery.
Therefore, the larger diagonal value $\omega_{max}=max\{\omega_{X},\omega_{Y}\}$ is used to normalize the coefficients of the $W$ matrix to make it always equivalent to a BS, that is,
\begin{equation}
\begin{split}
W^{0}&=W/\omega_{max}
=\begin{bmatrix}w_{XX}^{0}&w_{YX}^{0}\\w_{XY}^{0}&w_{YY}^{0}\end{bmatrix}\\&=U\Omega^{0}V=U\begin{bmatrix}\omega_{X}^{0}&0\\0&\omega_{Y}^{0}\end{bmatrix}V.
\end{split}
\end{equation}

The normalized Q-MIMO output is then given by
\begin{equation}
\begin{split}
   {S}_{out}^Q&=\sqrt{T}{\alpha}_{A}+U(\Omega^{0} V\hat{N}+\sqrt{1-{\omega^{0}_{X/Y}}^2}\hat{N}^{D})\\&=W^{0}\alpha_B+\hat{N}^{add}. 
\end{split}
\end{equation}
where $T$ is the transmittance between Alice and Bob and $\hat{N}^{add}=U\sqrt{1-{\omega^{0}_{X/Y}}^2}\hat{N}^{D}$.

Therefore, the Q-MIMO model can be described by physical devices within the EB model. 
Among them, the matrices $U$ and $V$ are naturally unitary, whose physical models are constructed by adjusting parameters such as BS ratio, phase modulator, and other passive devices. 
Instead, the non-unitary nature of the $W^{0}$ matrix will be fully represented by its singular value in the diagonal matrix $\omega^{0}_{X/Y}$, whose physical model corresponds to the BS model coupled with vacuum noise depending on the matrix element.

Given that the SVD decomposition itself is computationally intensive, we can directly derive the added noise $\hat{N}^{add}=[\hat{n}_X^{add},\hat{n}_Y^{add}]^T$
equivalent to the original MIMO output,
\begin{equation}
    \hat{n}_X^{add}=\sqrt{V_{XX}}\hat{n}_X;
\end{equation}
\begin{equation}
    \hat{n}_Y^{add}=\sqrt{V_{YY}-\frac{|V_{XY}|^2}{V_{XX}}}\hat{n}_Y+\frac{V_{XY}^*}{\sqrt{V_{XX}}}\hat{n}_X.
\end{equation}
Here, $\{\hat{n}_X,\hat{n}_Y\}$ stand for independent vacuum noise, 
$V_{XX}= 1-|w_{XX}^{0}|^2-|w_{YX}^{0}|^2$, $V_{YY} = 1-|w_{XY}^{0}|^2-|w_{YY}^{0}|^2$ and 
$V_{XY} = -{w_{XX}^{0}}{w_{XY}^{0}}^*-{w_{YX}^{0}}{w_{YY}^{0}}^*$.

\subsection{\label{sec:level2}Necessity of Q-MIMO shown by simulations}
To illustrate the necessity of Q-MIMO model, we explore a simulation with AWGN channel, which is usually used optical fiber channel.
Since the dynamic evolution process of 2*2 MIMO is related to the training sequence, eavesdroppers can control the channel condition including SOP and PDL, resulting the MIMO matrix non-unitary.
For the Gaussian channel in two polarization directions with parameters ${T_X,\epsilon_X}$ and ${T_Y,\epsilon_Y}$, the covariance matrix between the communicating parties in X-Pol can be directly written as
\begin{equation}
\gamma_{AB_X}=\begin{bmatrix}V_{A}I_2&\sqrt{T_X(V^2_A-1)}\sigma_z\\\sqrt{T_X(V^2_A-1)}\sigma_z&[T_X(V_A-1+\epsilon_X)+1]I_2\end{bmatrix},
\end{equation}
where $V_A$ is the variance of sending quantum state, $I_2$ is the identity matrix and $\sigma_z=\begin{bmatrix}1&0\\0&-1\end{bmatrix}$.

The non-unitary MIMO $W^{0} $ will result in the covariance matrix between the communicating parties becoming 
\begin{equation}\label{gamma}
\gamma_{AB_X}^{'}=\begin{bmatrix}V_{A}I_2&\sqrt{T^{'}_X(V^2_A-1)}\sigma_z\\\sqrt{T^{'}_X(V^2_A-1)}\sigma_z&[T^{'}_X(V_A-1+\epsilon_X)+1-V_{XX}]I_2\end{bmatrix}.
\end{equation}
\begin{figure}[ht]  %
    \centering  
    \includegraphics[width=0.42\textwidth]{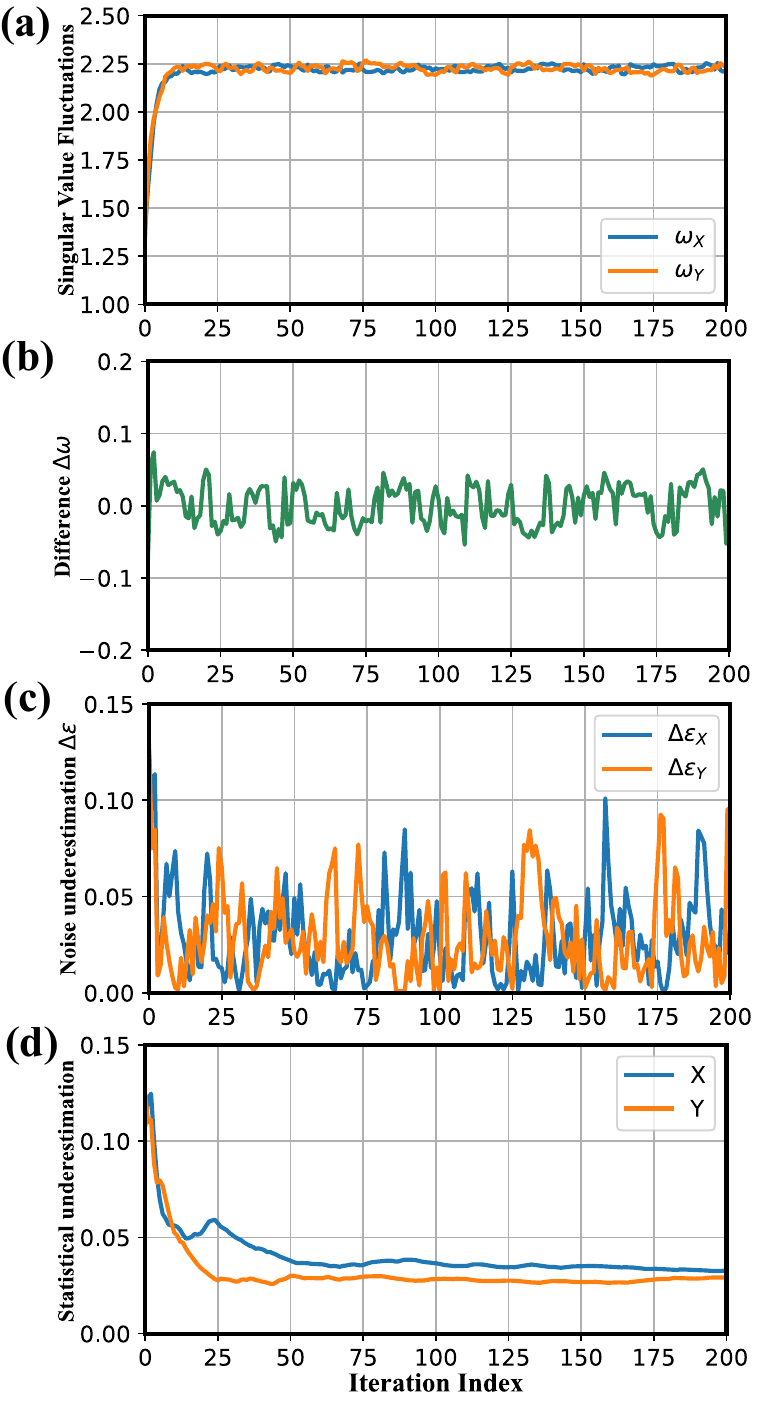}  
    \caption{(a) The singular value of the MIMO non-unitary in X-Pol and X-Pol. 
    (b) The difference curve between the two singular values of (a).
    (c) The real-time underestimation of excess noise.
    (d) The statistical integral average of excess noise variation of (c).
    }  
    \label{fig:simulations}  
\end{figure}
Here, $T'_X$ represents the overall X-pol transmittance after the dual-polarization signal passes through the $W$ matrix.

If the C-MIMO algorithm is not securely processed, the channel parameters of the system will be miscalculated.
It can be deduced that the estimated excess noise based on the MIMO output is
\begin{align}
\epsilon_X^e=\epsilon_X-\frac{(1-|w_{XX}^{0}|^2-|w_{YX}^{0}|^2)}{T_X^{'}}.
\end{align}
Similarly, the estimated result in Y-Pol is expressed as
\begin{align}
\epsilon_Y^e=\epsilon_Y-\frac{(1-|w_{XY}^{0}|^2-|w_{YY}^{0}|^2)}{T_Y^{'}}.
\end{align}
It can be seen that the non-unitary MIMO algorithm will cause an underestimation of the practical excess noise.

FIG. \ref{fig:simulations} illustrates the inherent characteristics of the dynamic MIMO algorithm and its impact on CV-QKD under simulated channel conditions, where the channel incorporates attenuation of 4dB and SOP with a rotation speed of 2krad/s.
(a) shows the evolution of the non-unitary property of the MIMO matrix in the two polarization directions with respect to the iterative index.
As the LMS algorithm updates, the equalizer coefficients gradually approach the optimal solution and the two singular values of the matrix fluctuate within a narrow range.
The difference between the two singular values $\Delta\omega=\omega_Y-\omega_X$ constitutes the origin of the dynamic MIMO security vulnerability, as demonstrated in (b). 
It can be observed that the singular value difference between the polarization directions fluctuates around zero, resulting in the normalized $W^{0}$ matrix consistently being non-unitary.
Clearly, it can be deduced that the presence of PDLs will amplify this discrepancy, leading to more severe security issues.
\begin{table}[t]
  \caption{\label{results}The parameter estimation results for C-MIMO and Q-MIMO model. PDL, polarization dependent loss; $T'$, the overall transmittance; $\epsilon^e$, the estimated excess noise.}
  \begin{ruledtabular}
  \begin{tabular}{ccccc}
    PDL & Polarization & $T'$ & $\epsilon^e$(SNU) & Model \\
    \midrule
    \multirow{4}{*}{0dB} & X & 0.395 & 0.117 & \multirow{2}{*}{C-MIMO} \\
    \cmidrule{2-4}
    & Y & 0.396 & 0.118 & \\
    \cmidrule(l){2-5}
    & X & 0.395 & 0.151 & \multirow{2}{*}{Q-MIMO} \\
    \cmidrule(l){2-4}
    & Y & 0.396 & 0.148 & \\
    \midrule
    \multirow{4}{*}{1dB} & X & 0.316 & -0.304\footnote{The incorrect negative excess noise occurs when using the standard C-MIMO algorithm without the proposed security corrections. This underestimation of excess noise will result in the generation of unsecure quantum keys.} & \multirow{2}{*}{C-MIMO} \\ 
    \cmidrule{2-4}
    & Y & 0.317 & -0.645$^{\text{a}}$ & \\ 
    \cmidrule{2-5}
    & X & 0.316 & 0.173 & \multirow{2}{*}{Q-MIMO} \\
    \cmidrule{2-4}
    & Y & 0.317 & 0.161 & \\
    \midrule
    \multirow{4}{*}{3dB} & X & 0.200 & -1.579$^{\text{a}}$ & \multirow{2}{*}{C-MIMO} \\ 
    \cmidrule{2-4}
    & Y & 0.201 & -2.889$^{\text{a}}$ & \\ 
    \cmidrule{2-5}
    & X & 0.200 & 0.248 & \multirow{2}{*}{Q-MIMO} \\
    \cmidrule{2-4}
    & Y & 0.200 & 0.202 & \\
  \end{tabular}
  \end{ruledtabular}
\end{table}
\begin{figure}[t]  %
    \centering  
    \includegraphics[width=0.42\textwidth]{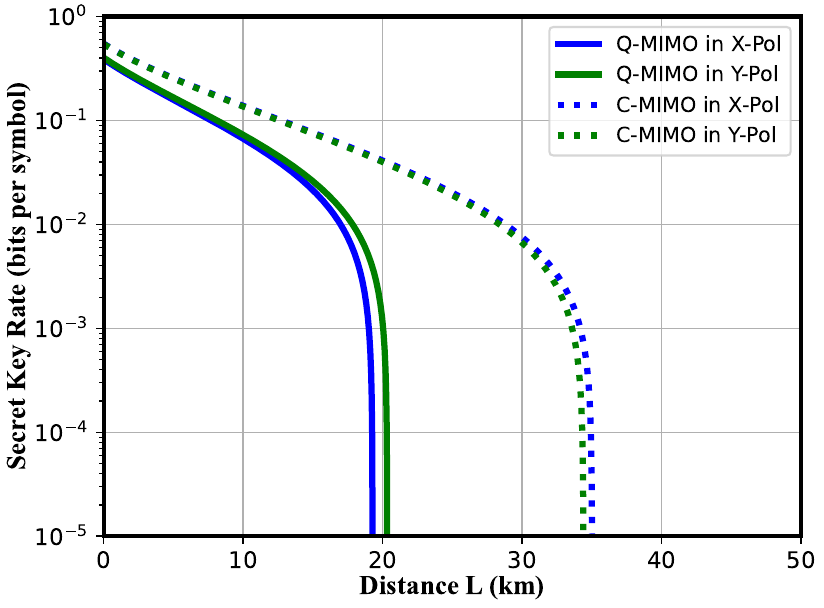}  
    \caption{Estimated secret key rate under two polarization directions in the simulated channel. The solid line represents Q-MIMO with $\epsilon_X=0.151$ SNU and $\epsilon_Y=0.148$ SNU, while the dashed line represents C-MIMO with $\epsilon_X=0.117$ SNU and $\epsilon_Y=0.118$ SNU. The simulation parameters are as follows: quantum state variance $V_A=4$, coordination efficiency $\beta=0.96$, and channel loss coefficient $\alpha=0.2$ dB/km.}  
    \label{fig:skr}  
\end{figure}
\begin{figure*}[t]  %
    \centering  
    \includegraphics[width=0.8\textwidth]{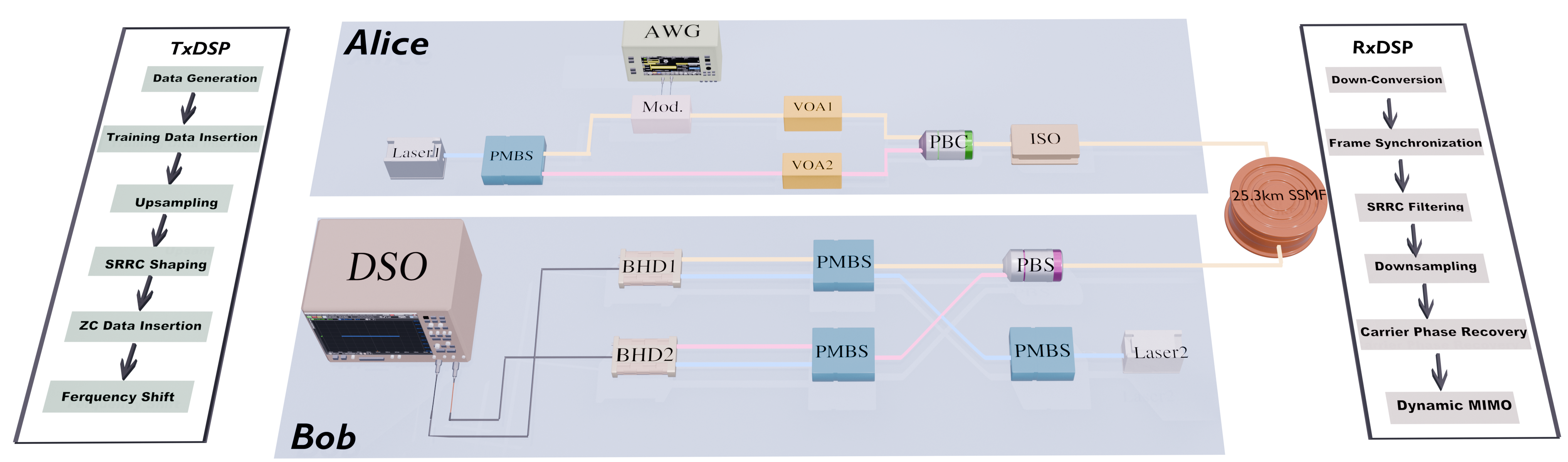}  
    \caption{Experimental setup of the LLO Gaussian-modulated CV-QKD system. Mod., IQ modulator; PMBS, polarization-maintaining beam splitter; VOA, variable optical attenuator; ISO, isolator; PBC, polarization beam combiner; SSMF, standard single-mode fiber.}  
    \label{fig:seup}  
\end{figure*}

(c) demonstrates the real-time underestimation of excess noise $\Delta \epsilon_X=\frac{(1-|w_{XX}^{0}|^2-|w_{YX}^{0}|^2)}{T_X'}$ and $\Delta \epsilon_Y=\frac{(1-|w_{XY}^{0}|^2-|w_{YY}^{0}|^2)}{T_Y'}$ by the MIMO matrices in Equations 23 and 24. 
Naturally, excess noise is preferable to be statistically estimated from a large amount of data, so we present the integral average of (c) in FIG. \ref{fig:simulations}(d).
The curve indicates that despite substantial real-time fluctuations, C-MIMO algorithms inevitably lead to underestimation of statistical excess noise and gradually enter a stable state.

Theoretical analysis has demonstrated the necessity of the secure Q-MIMO model for correct parameter estimation. 
TABLE.\ref{results} further presents the specific parameter estimation for C-MIMO and Q-MIMO.
It can be observed that C-MIMO algorithms will cause excess noise $\epsilon^e$ to be underestimated or even negative.
The results in the presence of channel PDL are also presented, where its existence leads to a larger difference $\Delta\omega$ and this security vulnerability increases as the PDL grows larger. 
The simulation of key rate without PDL is shown in FIG. \ref{fig:skr}, where the rate without trusted noise is significantly overestimated.
Consequently, the generation of unsecure keys demonstrates the necessity for security analysis of MIMO algorithm.
In practical systems where PDL exists, this security situation becomes even more severe.

\section{\label{sec:level1}Experimental Demonstration}
In this chapter, we shall conduct experimental verification of the Q-MIMO algorithm used for local local oscillator (LLO) CV-QKD system.
The specific experimental setup and corresponding experimental results are presented below.
\subsection{\label{sec:level2}Experimental setup}
The experimental setup for the digital CV-QKD system is shown in FIG. \ref{fig:seup}.
In the TxDSP module, the Gaussian-modulated symbols with a repetition frequency of 500 MHz are generated in a single polarization, which will be combined with the training sequence modulated by QPSK with 20dB higher power through time-division multiplexing.
The combined data are upsampled and then SRRC pulse-shaped with a roll-off factor of 0.3, where the Zadoff-Chu (ZC) sequence is added for subsequent frame synchronization.
Then, the frequency shift 1G is applied to the data to avoid low-frequency noise.

At the sending end, Alice's laser generates a continuous wave beam at 1550.11 nm with a linewidth of less than 100 Hz, which is then split into two directions by the PMBS.
The light in one polarization direction serves as the carrier for the quantum signal.
The preprocessed data are then modulated to the corresponding optical path by an IQ modulator, which is driven by 5GSa/s AWG with 16-bit resolution.
The light in another polarization direction is regarded as a single pilot for carrier recovery.
The two VOAs adjust their signals to the appropriate power, which are then merged through the PBC and transmitted into the 25.3 km fiber channel.

At the receiving end, we first use a PBS to split the received optical signal into two orthogonal polarization components (X-Pol and Y-Pol).
These two polarization components are then coherently mixed with LO beams whose frequencies are offset by 1.9 GHz from the signal carrier frequency.
The resulting intermediate-frequency (IF) electrical signals, which carry complete amplitude and phase information, are detected by the corresponding BPDs.
Finally, the receiver's output is sampled by the DSO with 10-bit resolution and fed into the Rx-DSP module.

The sampled IF signal is first digitally down-converted to recover four baseband in-phase/quadrature (x/p) components. 
Then, the ZC sequence achieves precise frame synchronization due to its excellent autocorrelation and cross-correlation characteristics. The synchronized data are filtered by the matched SRRC filter and downsampled to maximize the signal-to-noise ratio and suppress ISI. 
At this point, the pilot data are used to recover the random phase difference between the two independent lasers.
The SNU data undergo the static and deterministic process similar to quantum data to ensure security analysis.
The MIMO algorithm of LMS equalizer is employed for residual phase and polarization recovery, causing the data on the X-pol to converge towards the original quantum data from the transmitter, while suppressing the Y-pol data to zero.  
We individually perform the Q-MIMO model on the quantum data to ensure that the estimation of excess noise is fair and secure.
 
\subsection{\label{sec:level2}Experimental results}
\begin{figure}[t]  %
    \centering  
\includegraphics[width=0.4\textwidth]{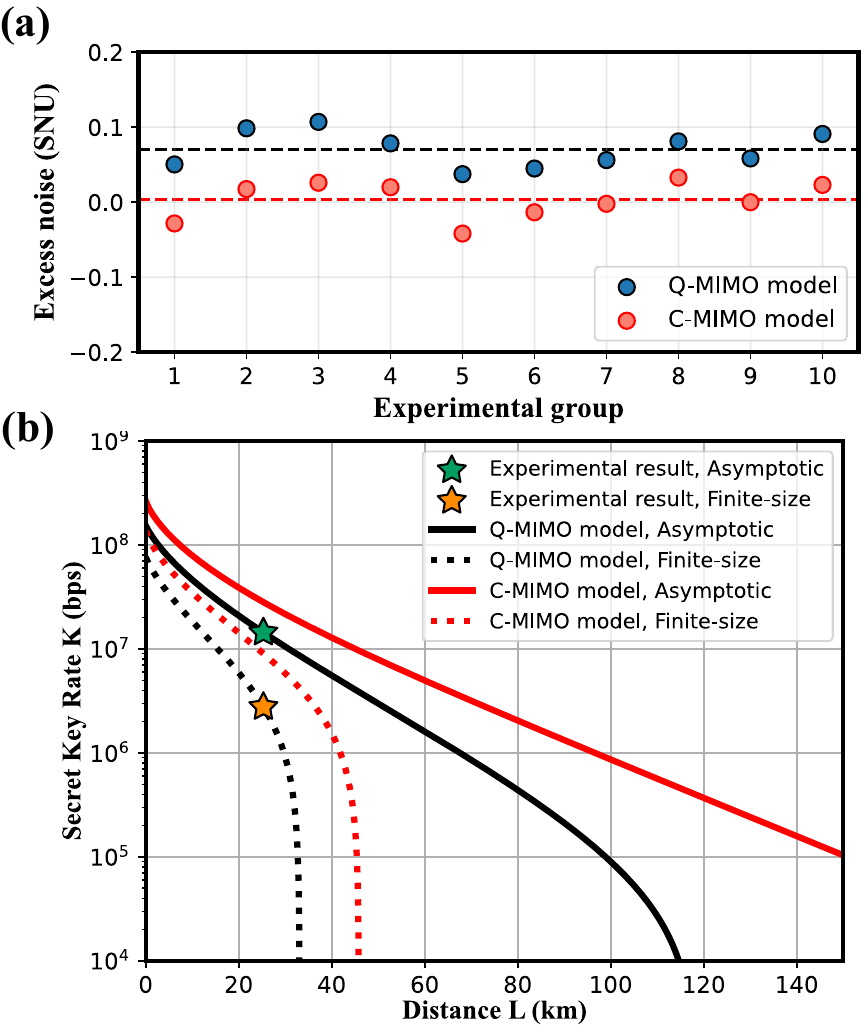}  
    \caption{
    (a) Estimated excess noise for ten groups of experimental data.
    The excess noise result of the blue sphere is estimated under the Q-MIMO model, with an average value of 0.07 SNU.
    Excess noise with C-MIMO are indicated by red spheres, with the average value underestimated to 0.008 SNU.
    (b) The secret key rate for C-MIMO and Q-MIMO model. The green star and yellow star represent the experimentally achieved secret key rate using the proposed Q-MIMO model in the asymptotical and finite-size case, respectively.
    The black curves are the simulation using experimental parameters with $\epsilon=0.07$ SNU in the asymptotical and finite-size case. 
    The red curves represent the incorrect key rate with $\epsilon=0.008$ SNU, where the original C-MIMO algorithm is applied.}  
    \label{fig:exper}  
\end{figure}
In the experiment, the modulation variance $V_A$ of quantum states at the transmitter end is set to 4 SNU. The trusted detector at the receiver end is pre-calibrated, with the detection efficiency $\eta_d=0.44$ and electrical nois $v_{el}=0.13$ SNU, respectively.
During the data acquisition and processing phase, we applied optimized DSP parameters and use proposed Q-MIMO model.
We have collected 10 groups of data sets, each with approximately $N=1\times 10^6$ quantum symbols, for precise channel parameter estimation.

From these data, we estimated the total channel transmittance $T_c=0.352$, corresponding to the expected fiber loss of 4.5 dB. 
10 blocks of estimated excess noise under the Q-MIMO algorithm are shown as blue spheres in the FIG. \ref{fig:exper} (a), with an average value of 0.07 SNU.
The estimated excess noise with C-MIMO is indicated by red spheres, even show negative values. 
The average across ten experimental groups is 0.008 SNU, far below the system's practical excess noise, which poses a serious security vulnerability.

Based on these parameters and the Gaussian channel assumption under optimal collective attacks, the asymptotic secret key rate is calculated using the standard formula 
\begin{equation}
	\label{K_{asy}}
	\begin{split}
		K_{asy}=f_{rep}(1-\alpha)[\beta I(A:B)-\chi(B:E)].
	\end{split}
\end{equation}
Here, $f_{rep}$ is the repetition frequency (or symbol rate) which determines the raw data generation rate, and $\alpha$ is the system overhead ratio accounting for non-key generation tasks such as dynamic MIMO training.
In addition, the Shannon mutual information between Alice and Bob is $I(A:B)$, while the Holevo bound $\chi(B:E)$ represents the theoretical information upper limit that an eavesdropper can acquire.
Reverse reconciliation is used for key generation, with a reconciliation efficiency of $\beta=0.96$.

In the finite-size case, the key rate formula is expressed as
\begin{equation}
	\label{K_{fin}}
	\begin{split}
		K_{fin}=f_{rep}(1-\alpha)\frac{n}{N}[\beta I(A:B)-S_{\epsilon_{PE}}(B:E)-\Delta(n)],
	\end{split}
\end{equation}
where $N$ is the total length of the data, $n$ of which is used to generate the key. We express the conditional entropy as $S_{\epsilon_{PE}}(B:E)$, it is the maximum value of Holevo information between Eve and Bob in the finite-size case with $\epsilon_{PE}=10^{-10}$.
The parameter $\Delta(N)$ is related to the security of private key amplification,
\begin{equation}
    \Delta(N)=(2dimH_X+3)\sqrt{\frac{log_2(2/\bar{\epsilon})}{n}}+\frac{2}{n}log_2(1/\epsilon_{PA}).
\end{equation}
where $H_X$ is the Hilbert space corresponding to the variable from Alice used in the raw key. $\bar{\epsilon}=10^{-10}$ is a smoothing parameter, and $\epsilon_{PA}=10^{-10}$ is the failure probability of the privacy key amplification procedure.

The final experimental results are presented in FIG. \ref{fig:exper} (b).
The solid black line represents the simulated secret key rate at the excess noise level of 0.07 SNU and the dashed one is the secret key rate in finite-size case.
At the transmission distance of 25.3 km, a secret key rate of 14.4 Mbps is achieved asymptotically and 2.76 Mbps is achieved in finite-size case, demonstrating the feasibility and effectiveness of our dynamic DSP security framework in a practical situation.
The red lines represent the key rate under underestimated noise in the asymptotical and finite-size case, where the insecure key distribution underscores the importance of proposed Q-MIMO modeling.

\section{\label{sec:level1}Discussion and conclusion}
\nocite{*}
In recent years, many high-performance CV-QKD systems have applied powerful DSP algorithms, which play a key role in digitally compensating for channel damage.
From the security perspective, continuous-mode field theory makes it possible to analyze its security. 
However, such rigorous security analysis remains limited to static linear algorithms, which are characterized by fixed transfer functions and can be modeled as unitary transformations.
In practice, the channel drift characteristics of dual-polarization system make dynamic MIMO algorithms essential for tracking and compensation.
The adaptive nature of the algorithm, where tap coefficients are adjusted in real-time based on the training sequence, means that we can no longer precisely describe its behavior using existing security framework.
This data-dependent training process introduces a potential side channel that may leak information to eavesdroppers, representing an important and unresolved challenge in the security proof of digital CV-QKD.

Therefore, this work focuses on secure modeling of dynamic DSP algorithms in the practical dual-polarization CV-QKD system. We conducted an in-depth analysis of the characteristics of the training C-MIMO matrix through SVD decomposition, and derived the trusted noise that needs to be added to correct the SNU.
The security of proposed Q-MIMO is fully proved and its necessity is presented through simulations.
Due to the presence of PDL and algorithmic constraints, the C-MIMO matrix will result in excess noise being underestimated, thereby generating unsecure keys.
Subsequently, we conducted a practical CV-QKD experiment with proposed Q-MIMO model, demonstrating that the secret rate can reach 14.4 Mbps@25.3km. 

By establishing the security framework for dynamic MIMO algorithms, we have pioneered a new paradigm for CV-QKD system design. 
This work will enable the complete replacement of cumbersome, inefficient and costly polarization compensation hardware with flexible, fast and cost-effective software solutions.
This software-defined algorithm not only reduces system complexity, but is also crucial for realizing next-generation high-capacity CV-QKD. 
The proposed Q-MIMO model provides the essential potential for dual-polarization systems, paving the way for system capacity to scale towards 100G/200G and beyond.

In summary, we have successfully established a theoretical security proof for a class of dynamic DSP algorithms, addressing the side-channel issue under MIMO equalization.
However, the path to secure digital CV-QKD does not limit to this, but should be extended to the entire DSP architecture. 
The digital down-conversion module implements heterodyne detection at the software level, whereas its security remains unexplored. Furthermore, our current work has not yet touched upon more complex multi-tap equalization and nonlinear DSP algorithms, whose complexity in processing temporal data may conceal unknown vulnerabilities.
These issues require future research to be completed, and only when a comprehensive and trustworthy DSP security framework is established can DSP be safely applied in practical CV-QKD systems.
\section{\label{sec:level1}ACKNOWLEDGMENTS}
This research was supported by the National Cryptologic Science Fund of China (2025NCSF02050) and the National Natural Science Foundation of China (U24B20135).
\bibliography{references}

\end{document}